\documentclass[10pt]{iopart} 
\usepackage[utf8]{inputenc}
\usepackage[UKenglish]{babel}
\usepackage{graphicx}
\usepackage{caption}
\usepackage{fullpage}
\usepackage{hyperref}
\usepackage{cleveref}
\usepackage{color}

\begin{document}
	
\title[$R_a$ as an estimate of $S_a$]{A simulated comparison between profile and areal surface parameters: $R_a$ as an estimate of $S_a$.}

\author{Henry T Lancashire}
\address{Aspire Centre for Rehabilitation Engineering and Assistive Technology, Department of Materials and Tissue, University College London, Stanmore, HA7 4LP, United Kingdom.}
\ead{henry.lancashire.10 [@] ucl.ac.uk}
\vspace{10pt}
\begin{indented}
\item[]\today
\end{indented}
\date{\today}

\begin{abstract} 

Direct comparison of areal and profile roughness measurement values is not advisable due to fundamental differences in the measurement techniques. However researchers may wish to compare between laboratories with differing equipment, or against literature values. This paper investigates how well the profile arithmetic mean average roughness, $R_a$, approximates its areal equivalent $S_a$.

Simulated rough surfaces and samples from the ETOPO1 global relief model were used. The mean of up to 20 $R_a$ profiles from the surface were compared with surface $S_a$ for 100 repeats.

Differences between $\bar{R_a}$ and $S_a$ fell as the number of $R_a$ values averaged increased. For simulated surfaces mean \% difference between $\bar{R_a}$ and $S_a$ was in the range 16.06\% to 3.47\% when only one $R_a$ profile was taken. By averaging 20 $R_a$ values mean \% difference fell to 6.60\% to 0.81\%. By not considering $R_a$ profiles parallel to the main feature direction (identified visually), mean \% difference was further reduced. For ETOPO1 global relief surfaces mean \% difference was in the range 52.09\% to 22.60\%  when only one $R_a$ value was used, and was 33.22\% to 9.90\% when 20 $R_a$ values were averaged. Where a surface feature direction could be identified, accounting for reduced the difference between $\bar{R_a}$ and $S_a$ by approximately 5\% points.
  
The results suggest that taking the mean of between 3 and 5 $R_a$ values will give a good estimate of $S_a$ on regular or simple surfaces. However, for some complex real world surfaces discrepancy between $\bar{R_a}$ and $S_a$ are high. Caveats including the use of filters for areal and profile measurements, and profile alignment are discussed.
   
\end{abstract}



\vspace{2pc}



\section{Introduction}

Surface roughness is often reported as a parameter of a profile, along 2-dimensional line \cite{gadelmawla_roughness_2002}. Recently focus has shifted to using areal parameters, in 3-dimensions. Areal roughness parameters may better describe surfaces, in particular when anisotropic features, such as grooves are present on the surface \cite{jiang_paradigm_2007,leach_guide_2008}. 

Areal measurements can be made using stylus systems, similarly to profile measurements, and also using optical and atomic force instruments \cite{jiang_paradigm_2007,leach_introduction_2011,galgoczy_atomic_2001}. Measurement of the areal equivalents of the parameters of a profile may not always be possible due to equipment constraints. In addition, there is a need to compare recent areal measurements against literature values measured using profile techniques. Therefore the extent to which profile parameters approximate areal parameters should be assessed on a variety of surfaces.

\begin{figure*}[htbp]
\centering
\includegraphics[width=0.5\textwidth]{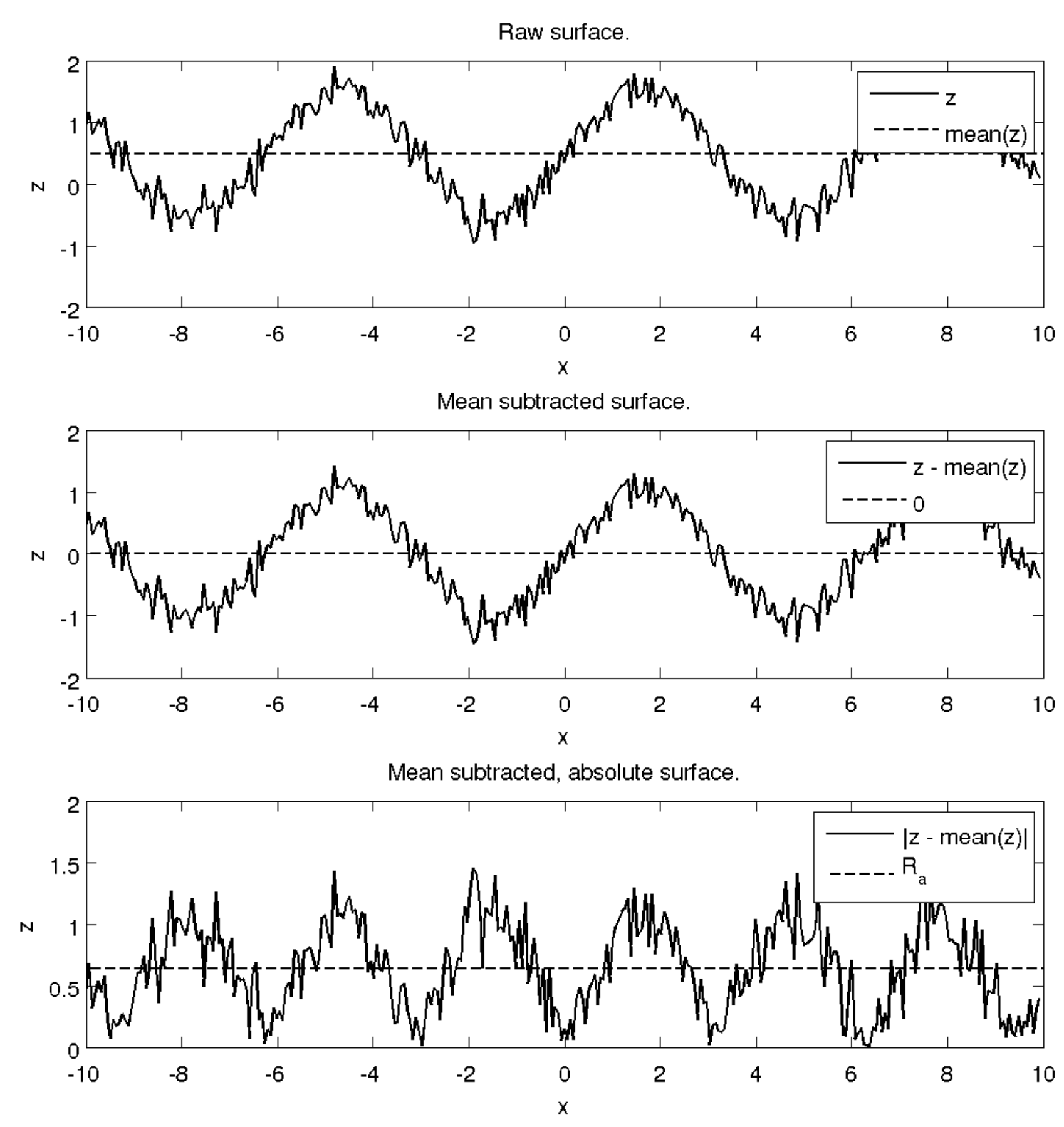}
\caption{Calculating $R_a$ from a surface as described in equation \ref{eq:Ra2}. The mean of the raw surface is calculated (top). The mean is subtracted from all surface points (middle). The absolute value of all mean subtracted surface points is taken (bottom). The mean of these absolute deviations from the mean surface height is the $R_a$ (bottom).}
\label{fig:Ra}
\end{figure*}

This paper will focus on the $R_a$ parameter (equations \ref{eq:Ra} and \ref{eq:Ra2}) and its areal equivalent, $S_a$ (equations \ref{eq:Sa} and \ref{eq:Sa2}) \cite{iso_25178,iso_4287,leach_fundamental_2010}. Both parameters are the arithmetic mean average absolute deviation from the mean height of the surface, but have some fundamental differences due to the use of profile or areal filters \cite{leach_fundamental_2010}. Alternative measures which better characterise and descriminate surfaces can be used \cite{jiang_paradigm_2007,bigerelle_assessing_2017}. However, $R_a$ is the most widely reported parameter, used by over 60\% of respondents in a 2003 to 2004 industrial consultation \cite{blunt_development_2008}. Figure \ref{fig:Ra} shows the $R_a$ algorithm visually.

\numparts
\begin{eqnarray}
R_{a} = \frac{1}{l}  \int_0^l |Z(x)|{\rm d}x \label{eq:Ra}\\
R_{a} = \frac{1}{N} \sum_{i=0}^{N} |Z_{i}| \qquad{N\ {\rm is\ the\ number\ of\ points\ in\ } x.} \label{eq:Ra2}
\end{eqnarray}
\endnumparts

\numparts
\begin{eqnarray}
S_{a} = \frac{1}{A} \int\int_{A} |Z(x,y)|{\rm d}x{\rm d}y \label{eq:Sa}\\
S_{a} = \frac{1}{MN} \sum_{j=1}^{N}\sum_{i=1}^{M} |Z(x_{i},y_{j})| \qquad{{\rm (}M,\ N{\rm )\ are\ the\ number\ of\ points\ in\ (}x,\ y{ \rm).}} \label{eq:Sa2}
\end{eqnarray}
\endnumparts

On a perfectly flat surface $S_a = R_a = 0$. However, patterned surfaces can be readily conceived which have different $R_a$ and $S_a$ values. For example, the surface defined by equation \ref{eq:6} (see figure \ref{fig:surfaces}). In this case $S_a$ is non-zero, while any profile taken at constant $x$ will have $R_a = 0$. Therefore a single $R_a$ value is not a valid estimate of $S_a$.

It is proposed that the mean average of a number of $R_a$ values ($\bar{R_a}$) will better approximate $S_a$. This paper will investigate to what extent the average $\bar{R_a}$ from a finite number of profiles approximates $S_a$. In addition, this paper will investigate the role of $R_a$ profile direction when estimating $S_a$.

\section{Methods}

\subsection{Simulated Surface Topography}

\begin{figure*}[htbp]
\centering
\includegraphics[width=1.00\textwidth]{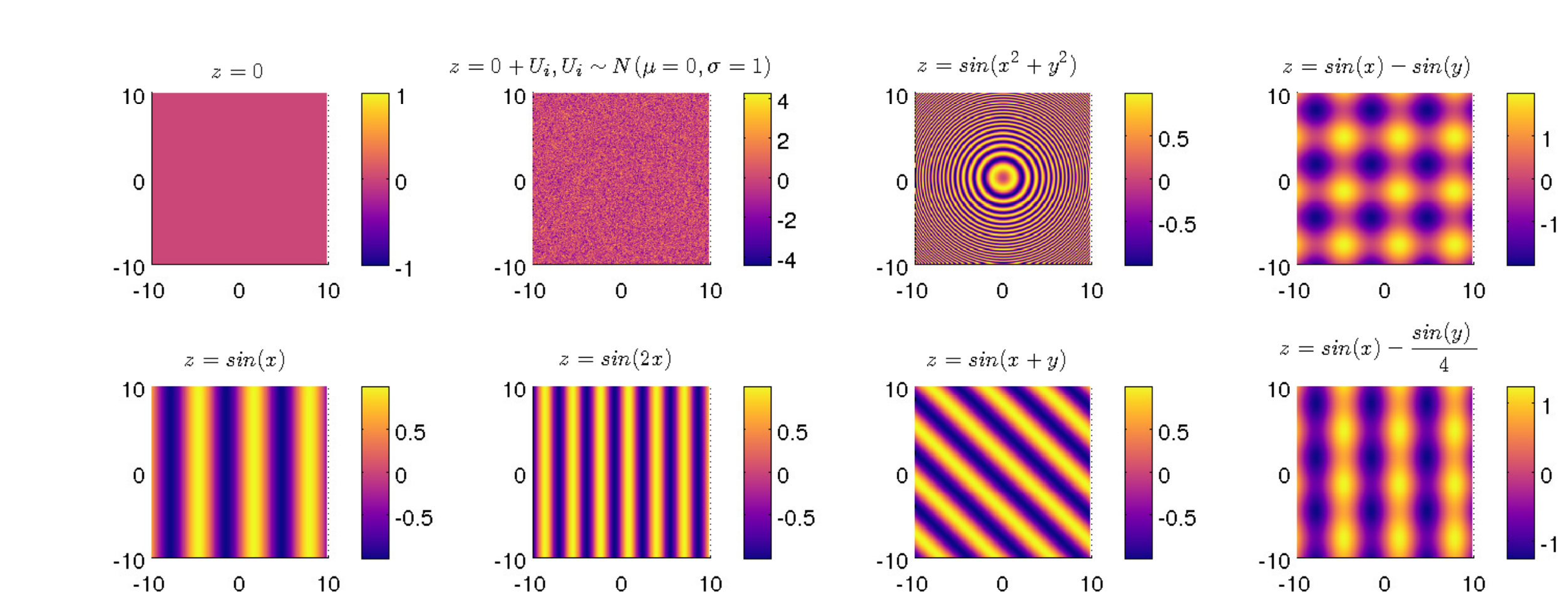}
\caption{Simulated underlying surface topographies. Additional roughness was added to each surface, as shown by $z = 0 + U_{i}, U_{i} \sim N({\mu}=0,{\sigma}=1)$.}
\label{fig:surfaces}
\end{figure*}

Simulations were caried out using {MATLAB} (2013a, The Mathworks Inc., Natick, USA, on Linux 64-bit). Surfaces were modelled in the range -10 $\leq$ ($x,y$) $\leq$ 10 on a 100,000 point mesh. The following surfaces were generated: a planar surface, equation \ref{eq:1}; a ring shaped surface, equation \ref{eq:2}; a simple undulating surface, equation \ref{eq:3}; wide vertical ridges, equation \ref{eq:6}; narrow vertical ridges, equation \ref{eq:7}; diagonal ridges, equation \ref{eq:5}; a complex undulating surface, equation \ref{eq:4}. Surface plots are shown in figure \ref{fig:surfaces}.
\begin{eqnarray} 
z = 0 \label{eq:1}\\
z = sin(x^{2} + y^{2}) \label{eq:2}\\
z = sin(x) - sin(y) \label{eq:3}\\
z = sin(x) \label{eq:6}\\
z = sin(2x) \label{eq:7} \\
z = sin(x+y) \label{eq:5}\\
z = sin(x) - \frac{sin(y)}{4} \label{eq:4}
\end{eqnarray}

Normally distributed random roughness ($\mu$ = 0, $\sigma$ = 1) was added to each surface to create new surfaces for each simulation. $S_a$ was calculated for the whole surface and $R_a$ was calculated for vertical (x=0), horizontal (y=0), 45$^{\circ}$ and -45$^{\circ}$ sweeps across the surface. $R_a$ values were calculated for every vertical and horizontal in the 100,000 point mesh, and $R_a$ values were calculated from the 45$^{\circ}$ and -45$^{\circ}$ diagonals and antidiagonals respectively, only diagonals from -158 to 158 were considered, where 1 is the main diagonal or antidiagonal.  For each surface from 1 to 20 $R_a$ values are chosen at random and an average is taken ($\bar{R_a}$), each $\bar{R_a}$  is calculated with a new randomly generated roughness. This is repeated n = 100 times.

Where the surface has a distinct directional texture, in addition to calculating $R_a$ values in the four directions described above, simulations were run where no $R_a$ values were calculated parallel to the main features. For example with vertical ridges, simulations were also run without $R_a$ values calculated from vertical sweeps.

\subsection{Earth Surface Topography}

Comparisons were also made using publicly available topography data. Ten random samples from the ETOPO1 global relief model of Earth were taken using the bedrock dataset \cite{amante_etopo1_2009}. Each sample was a 10$^{\circ}$ by 10$^{\circ}$ mesh of 360,000 points. As with the simulated surfaces $S_a$ was calculated for the whole surface and $R_a$ was calculated for vertical, horizontal, 45$^{\circ}$ and -45$^{\circ}$ sweeps across the surface as above. The means of t $R_a$ values were calculated for t = 1 to t = 20. This was repeated 100 times for each of the 10 ETOPO1 samples.

If a distinct directional texture could be identified visually calculations were also run without $R_a$ values from sweeps parallel to the main features.

\subsection{Data Analysis}

The absolute value of the percentage difference between $\bar{R_a}$ and $S_a$ was calculated for each mean and surface. Data are plotted as mean $\pm$ 95\% confidence intervals of the of absolute percentage difference to show how well $\bar{R_a}$ can approximate $S_a$. Data are reported as mean $\pm$ standard devation.

\section{Results}

\subsection{Simulated Surface Topography}

\begin{figure*}[htbp]
\centering
\includegraphics[width=1.00\textwidth]{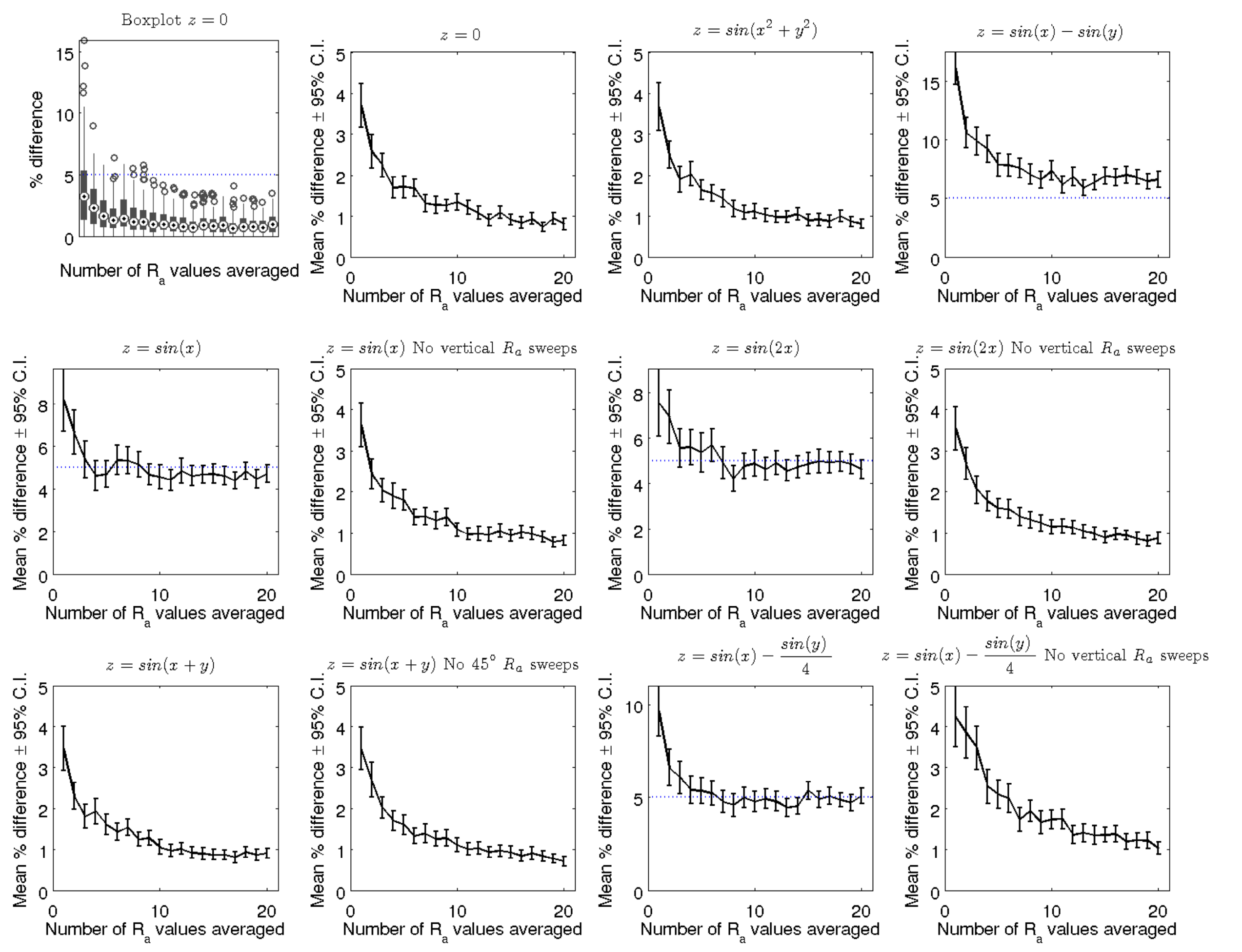}
\caption{Charts of the extent to which $\bar{R_a}$ approximates $S_a$ on simulated surfaces with random roughness based upon a range of underlying surfaces. A boxplot of results from a flat ($z = 0$) surface is given to show the distribution of \% differences. All other plots show the absolute value of the mean \% difference ($\pm$95\% C.I.) between $\bar{R_a}$ and $S_a$ for the given surface plotted against the number of sweeps used to calculated $\bar{R_a}$. For surfaces with distinct directional texture (for example vertical ridges, $z = sin(x)$) results are also reported for calculations without $R_a$ in parallel to the main features (i.e. no vertical $R_a$ profiles are used for $z = sin(x)$). Between 1 and 20 $R_a$ values are used to calculate $\bar{R_a}$, n = 100 surfaces with random roughness are generated to create each data point.}
\label{fig:outputs}
\end{figure*}

$R_a$ and $S_a$ values were successfully calculated for the surfaces considered. Results for the simulated surfaces are shown in figure \ref{fig:outputs}, and values are reported in table \ref{tab:outputs}. For all surfaces mean \% difference between $\bar{R_a}$ and $S_a$ decreases with increasing number of $R_a$ averaged. However, for some surfaces a plateau is observed after a finite number of $R_a$ values averaged. 

Plateaus in mean \% difference are present in the data for surfaces with distinct directional features, for example the vertical ridges of $z = sin(x)$. Where the mean \% difference between $\bar{R_a}$ and $S_a$ was calculated without $R_a$ profiles taken parallel to the main features and there was a drop in \% difference for all measurements: for vertical ridges ($z = sin(x)$) \% difference was reduced from 8.18$\pm$7.54\% to 3.63$\pm$2.69\% when only taking one $R_a$ profile as an $S_a$ estimator (when 20 $R_a$ profiles were averaged \% difference was reduced from 4.72$\pm$2.15\% to 0.823$\pm$0.617\%. The reduction in \% difference was small for the diagonal ridges, $z = sin(x + y)$ (3.47$\pm$2.73\% to 3.47$\pm$2.63, and 0.909$\pm$0.618\% to 0.712$\pm$0.610\%, for 1 and 20 $R_a$ profiles averaged respectively).

\fulltable{\label{tab:outputs}Table of \% difference between $\bar{R_a}$ and $S_a$ with between 1 and 20 $R_a$ values averaged, mean $\pm$ standard devation for results from 100 surfaces using the equation reported and additional random roughness.
For clarity and space some values are omitted.}
\br
& \centre{7}{\% difference between $\bar{R_a}$ and $S_a$ by number of $R_a$ values averaged} & \\ \ns
Surface & \crule{7} & \\
Equation & 1 & 2 & 3 \ldots & 5 \ldots & 10 \ldots & 15 \ldots & 20 & Notes \\
\mr
$z = 0$ & \03.69$\pm$2.73 & \02.58$\pm$2.08 & 2.25$\pm$1.48 & 1.71$\pm$1.39 & 1.34$\pm$1.00 & 0.91$\pm$0.71 & 0.81$\pm$0.71 & $^{\rm a}$ \\
$z = sin(x^{2} + y^{2})$ & \03.67$\pm$2.96 & \02.50$\pm$1.66 & 1.90$\pm$1.59 & 1.63$\pm$1.23 & 1.12$\pm$0.89 & 0.90$\pm$0.69 & 0.81$\pm$0.60 & $^{\rm a}$ \\
$z = sin(x) - sin(y)$ & 16.06$\pm$6.95 & 10.59$\pm$6.57 & 9.85$\pm$5.89 & 7.87$\pm$4.90 & 7.35$\pm$4.31 & 6.91$\pm$3.54 & 6.60$\pm$3.45 & $^{\rm a}$ \\
$z = sin(x)$ & \08.18$\pm$7.54 & \06.65$\pm$5.26 & 5.38$\pm$4.36 & 4.69$\pm$3.21 & 4.56$\pm$2.86 & 4.71$\pm$2.35 & 4.72$\pm$2.15 & $^{\rm a}$ \\
$z = sin(x)$ & \03.63$\pm$2.69 & \02.44$\pm$1.87 & 2.04$\pm$1.44 & 1.80$\pm$1.29 & 1.08$\pm$0.79 & 0.94$\pm$0.73 & 0.82$\pm$0.62 & $^{\rm b}$ \\
$z = sin(2x)$ & \07.55$\pm$7.57 & \06.91$\pm$6.05 & 5.54$\pm$4.23 & 5.35$\pm$4.39 & 4.87$\pm$2.89 & 4.87$\pm$2.39 & 4.61$\pm$2.01 & $^{\rm a}$ \\
$z = sin(2x)$ & \03.54$\pm$2.71 & \02.69$\pm$1.97 & 2.05$\pm$1.71 & 1.59$\pm$1.17 & 1.15$\pm$0.88 & 0.90$\pm$0.67 & 0.88$\pm$0.69 & $^{\rm b}$ \\
$z = sin(x+y)$ & \03.47$\pm$2.73 & \02.30$\pm$1.67 & 1.80$\pm$1.47 & 1.61$\pm$1.25 & 1.06$\pm$0.90 & 0.87$\pm$0.68 & 0.91$\pm$0.62 & $^{\rm a}$ \\
$z = sin(x+y)$ & \03.47$\pm$2.63 & \02.69$\pm$2.17 & 2.01$\pm$1.33 & 1.60$\pm$1.24 & 1.11$\pm$0.86 & 0.93$\pm$0.77 & 0.71$\pm$0.61 & $^{\rm c}$ \\
$z = sin(x) - \frac{sin(y)}{4}$ & \09.65$\pm$6.95 & \06.60$\pm$4.94 & 6.06$\pm$4.47 & 5.35$\pm$3.54 & 4.76$\pm$2.84 & 5.36$\pm$2.42 & 5.06$\pm$2.06 & $^{\rm a}$ \\
$z = sin(x) - \frac{sin(y)}{4}$ & \04.24$\pm$3.79 & \03.85$\pm$3.15 & 3.47$\pm$2.66 & 2.34$\pm$1.77 & 1.73$\pm$1.14 & 1.36$\pm$0.93 & 1.03$\pm$0.78 & $^{\rm b}$ \\
\br
\end{tabular*}
\noindent $^{\rm a}$ $R_a$ profile sweeps taken at 0$^{\circ}$, 45$^{\circ}$, -45$^{\circ}$, and 90$^{\circ}$.
\noindent $^{\rm b}$ $R_a$ profile sweeps taken at 0$^{\circ}$, 45$^{\circ}$, and -45$^{\circ}$. No $R_a$ calculated parallel to the main features.
\noindent $^{\rm c}$ $R_a$ profile sweeps taken at 0$^{\circ}$, 45$^{\circ}$, and 90$^{\circ}$. No $R_a$ calculated parallel to the main features.
\end{table}

\subsection{Earth Surface Topography}

\begin{figure*}[htbp]
\centering
\includegraphics[width=1.00\textwidth]{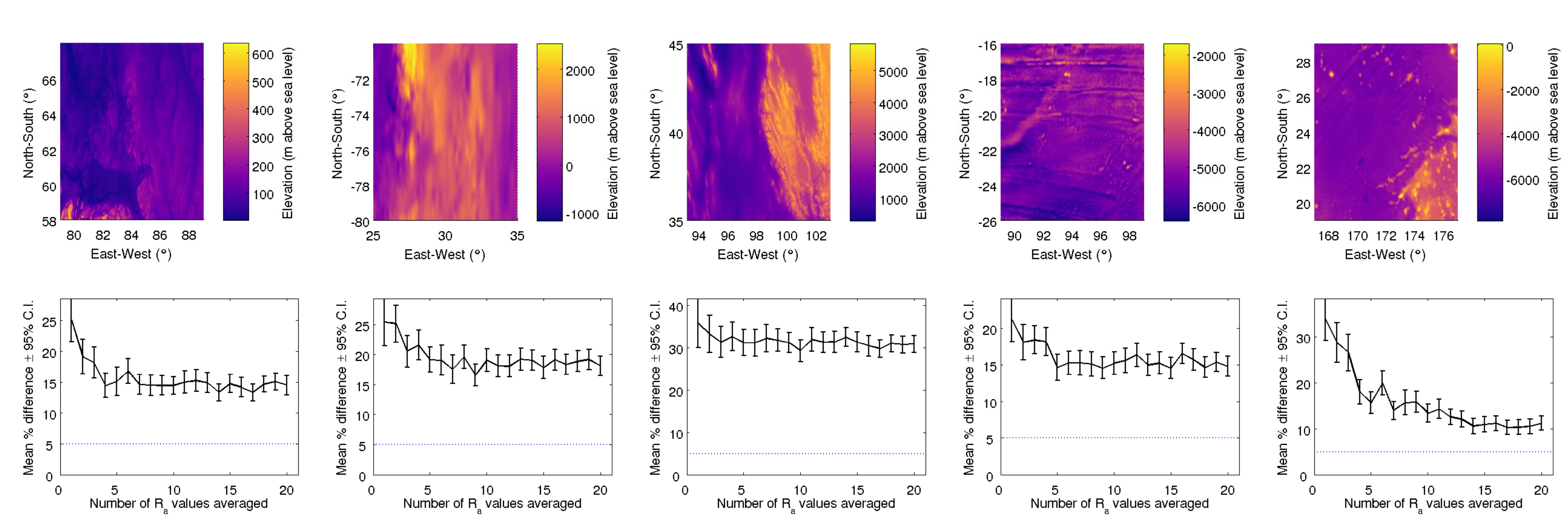}
\caption{Charts of the extent to which $\bar{R_a}$ approximates $S_a$ on the ETOPO1 global relief model \cite{amante_etopo1_2009}. Charts correspond to the above surface sample.} 
\label{fig:ETOPO_outputs}
\end{figure*}

Analysis of samples from the ETOPO1 global relief model \cite{amante_etopo1_2009} showed that mean \% difference between $\bar{R_a}$ and $S_a$ decreases with increasing number of $R_a$ averaged, see figure \ref{fig:ETOPO_outputs} and table \ref{tab:ETOPO_outputs}. As for the simulated surfaces, for some ETOPO1 samples a plateau was observed after which increasing the number of $R_a$ averaged did not appear to improve mean \% difference.

Overall \% differences were larger for the Earth surface samples than for the simulated surfaces, with mean differences in the ranges 52.09\% to 22.60\% and 16.06\% to 3.47\% respectively when one $R_a$ value was used (and 33.22\% to 9.90\% and 6.60\% to 0.81\% respectively when 20 $R_a$ values are used). 

Where there was a distinct directional texture this could be accounted for in the analysis. Improvement in mean \% difference was observed for samples 2, 4, and 5 when $R_a$ profiles parallel to the main feature direction were not included (table \ref{tab:ETOPO_outputs}).

\fulltable{\label{tab:ETOPO_outputs}Table of \% difference between $\bar{R_a}$ and $S_a$ with between 1 and 20 $R_a$ values averaged, mean $\pm$ standard devation for results from the ETOPO1 global relief model with 10$^{\circ}$ by 10$^{\circ}$ samples centred on the reported points. Values are calculated from 100 repetitions using randomly chosen $R_a$ profiles.
For clarity and space some values are omitted.}
\br
& \centre{7}{\% difference between $\bar{R_a}$ and $S_a$ by number of $R_a$ values averaged} & \\ \ns
Surface & \crule{7} & \\
Sample & 1 & 2 & 3 \ldots & 5 \ldots & 10 \ldots & 15 \ldots & 20 & Notes \\
\mr
1 \0(63$^{\circ}$N \084$^{\circ}$E) & 23.67$\pm$17.01 & 20.11$\pm$13.51 & 18.25$\pm$12.27 & 15.14$\pm$10.37 & 14.54$\pm$\08.79 & 13.87$\pm$\07.06 & 13.59$\pm$\06.40 & $^{\rm a}$  \\
2 \0(75$^{\circ}$S \030$^{\circ}$E) & 29.57$\pm$25.66 & 23.78$\pm$17.50 & 19.20$\pm$12.15 & 20.79$\pm$12.17 & 19.16$\pm$\09.28 & 19.16$\pm$\08.34 & 18.27$\pm$\07.82 & $^{\rm a}$  \\
2 \0(75$^{\circ}$S \030$^{\circ}$E) & 20.70$\pm$16.04 & 14.72$\pm$10.41 & 14.84$\pm$10.30 & 13.09$\pm$\08.08 & 10.28$\pm$\06.99 & 12.04$\pm$\06.95 & 10.66$\pm$\06.25 & $^{\rm b}$ \\
3 \0(22$^{\circ}$N \055$^{\circ}$E) & 52.09$\pm$32.59 & 44.22$\pm$25.91 & 40.22$\pm$24.62 & 34.30$\pm$19.48 & 29.08$\pm$17.60 & 26.42$\pm$14.26 & 27.51$\pm$15.32 & $^{\rm a}$ \\
4 \0(40$^{\circ}$N \098$^{\circ}$E) & 29.86$\pm$27.17 & 35.25$\pm$24.08 & 31.49$\pm$20.75 & 29.82$\pm$15.24 & 32.33$\pm$12.67 & 31.84$\pm$11.48 & 33.22$\pm$11.13 & $^{\rm a}$ \\
4 \0(40$^{\circ}$N \098$^{\circ}$E) & 34.57$\pm$29.41 & 25.36$\pm$19.11 & 28.51$\pm$19.11 & 26.52$\pm$15.85 & 23.53$\pm$13.51 & 25.28$\pm$13.89 & 25.02$\pm$13.49 & $^{\rm b}$ \\
5 \0(76$^{\circ}$S \052$^{\circ}$E) & 24.52$\pm$19.65 & 23.24$\pm$14.47 & 22.02$\pm$12.17 & 21.41$\pm$11.52 & 22.19$\pm$\09.87 & 20.63$\pm$\07.96 & 22.46$\pm$\08.11 & $^{\rm a}$ \\
5 \0(76$^{\circ}$S \052$^{\circ}$E) & 22.50$\pm$13.66 & 17.09$\pm$11.19 & 15.51$\pm$\09.83 & 15.12$\pm$\09.06 & 14.83$\pm$\06.75 & 16.16$\pm$\05.37 & 16.35$\pm$\05.85 & $^{\rm b}$ \\
6 \0(21$^{\circ}$S \094$^{\circ}$E) & 22.60$\pm$12.78 & 19.39$\pm$13.04 & 17.36$\pm$10.05 & 17.08$\pm$\08.87 & 16.64$\pm$\08.66 & 15.99$\pm$\06.35 & 15.77$\pm$\07.21 & $^{\rm a}$ \\
7 \0(61$^{\circ}$S \031$^{\circ}$E) & 43.69$\pm$33.16 & 31.58$\pm$26.16 & 33.10$\pm$24.07 & 29.44$\pm$15.86 & 23.09$\pm$11.86 & 24.23$\pm$11.39 & 23.23$\pm$10.19 & $^{\rm a}$ \\
8 \0(44$^{\circ}$S 165$^{\circ}$E) & 33.90$\pm$21.52 & 23.55$\pm$16.28 & 19.68$\pm$14.79 & 19.68$\pm$13.82 & 18.24$\pm$12.73 & 17.18$\pm$10.51 & 17.79$\pm$10.23 & $^{\rm a}$ \\
9 \0(24$^{\circ}$N 172$^{\circ}$E) & 39.82$\pm$30.39 & 28.13$\pm$21.14 & 25.81$\pm$18.02 & 18.01$\pm$13.11 & 12.59$\pm$10.61 & 10.74$\pm$\07.96 & \09.90$\pm$\07.30 & $^{\rm a}$  \\
10 (64$^{\circ}$N \032$^{\circ}$E) & 24.26$\pm$17.53 & 24.59$\pm$16.04 & 23.58$\pm$13.30 & 20.25$\pm$10.55 & 19.00$\pm$\08.75 & 18.93$\pm$\07.79 & 18.99$\pm$\07.12 & $^{\rm a}$  \\
\br
\end{tabular*}
\noindent $^{\rm a}$ $R_a$ profile sweeps taken at 0$^{\circ}$, 45$^{\circ}$, -45$^{\circ}$, and 90$^{\circ}$.
\noindent $^{\rm b}$ $R_a$ profile sweeps taken at 0$^{\circ}$, 45$^{\circ}$, and -45$^{\circ}$. No $R_a$ calculated parallel to the main feature(s).
\end{table}

\section{Discussion}

The results show that for a given surface the average of a finite number of $R_a$ values can approximate $S_a$. Results from simulated surfaces show that when accounting for the main feature direction (for example grooves), only taking $R_a$ profiles at greater than 45$^{\circ}$ from features (not parallel),  mean \% difference less than 5\% is achievable when averaging 3 random $R_a$ profiles. 

The exception for the simulated surfaces is $z = sin(x) - sin(y)$, with regular peaks and pits, compared with ridges and grooves on other surfaces (figure \ref{fig:surfaces}), this may be because very few of the randomly selected $R_a$ profiles capture the full extent of the height deviation: vertical and horizontal can only capture peaks and saddles, or troughs and saddles, not both peaks and troughs in the same sweep. Altering the range of angles used to capture $R_a$ may impro 

Results from Earth surface samples show that the \% differences achievable on simulated surfaces may not be achievable in real world applications: mean \% difference between 40.22\% and 17.36\% are achievable when averaging 3 random $R_a$ profiles, and accounting for the main feature direction reduces mean \% differences by circa 5\% points.

Caveats on the direct comparison of $R_a$ and $S_a$ are given by Leach, 2010 \cite{leach_fundamental_2010}. In this study both $\bar{R_a}$ and $S_a$ were calculated over the same distance or area; however, if the lengths or areas are different this will influence the reported values, especially on surfaces with multi-scale features \cite{jiang_paradigm_2007}. Leach comments that $R_a$ should be measured over ``a number of consecutive sampling lengths'' with users reporting the average value \cite{leach_fundamental_2010}. 

\begin{figure*}[htbp]
\centering
\includegraphics[width=1\textwidth]{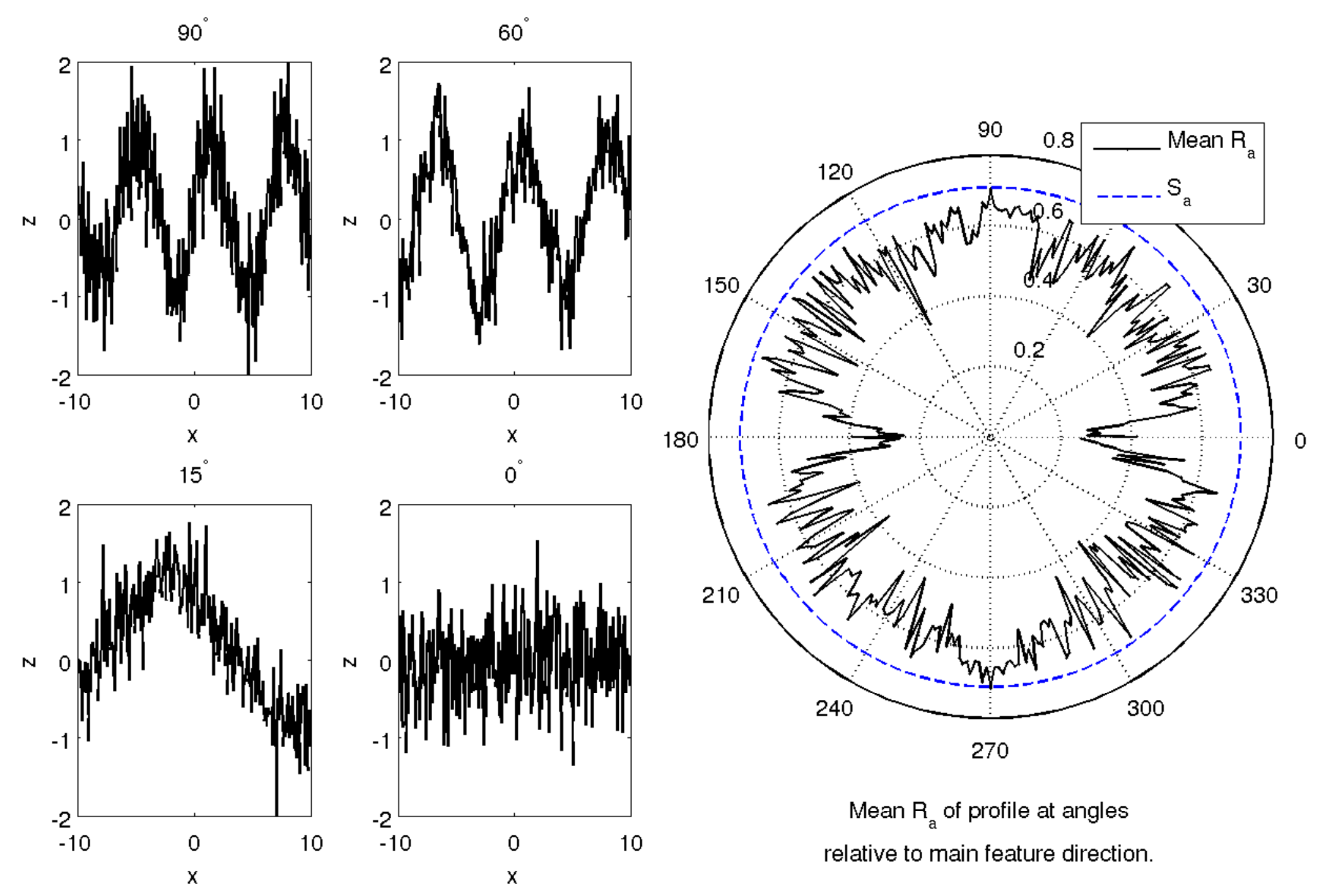}
\caption{The effect of rotation on $\bar{R_{a}}$ for wide vertical ridges, equation \ref{eq:6}, wide vertical ridges, with added random roughness $N(\mu = 0, \sigma = 0.5)$, $\sigma = 0.5$ used for clarity. At 0$^{\circ}$ and 180$^{\circ}$ $R_a$ profiles were taken parallel to the main feature direction. A reduction in $\bar{R_{a}}$ is observed for $R_a$ profiles parallel to the main features. An increased gradient and greater number of peaks is observed when profiles are taken perpendicular to the main features (compared to non-perpendicular profiles). $\bar{R_{a}}$ is consistently close to $S_a$ when profiles are taken near the perpendicular (90$^{\circ}$ and 270$^{\circ}$). $\bar{R_{a}}$ values are the mean of 3 $R_a$ values.}
\label{fig:Rotation}
\end{figure*}

Improving the accuracy of $\bar{R_a}$ as a representation of $S_a$ can depend upon knowing the main feature direction, and taking random profiles $\leq \pm$45$^{\circ}$ from perpendicular to this direction. Determining main feature direction is possible by eye or with low magnification on many machine surfaces, however this may not be possible with nanoscale features. If only profilometry is available, and nothing is known about the surface texture examining sweeps at equally spaced angles (from 0$^{\circ}$ to 90$^{\circ}$) can give an understanding of texture direction, where sweeps parallel to the main features will have shallower features with the smallest feature gradients, while sweeps perpendicular to the main features will be deeper and have the greatest gradients. Care must be taken to determine whether features seen on profiles represent pits / peaks, which do not have directionality, and ridges / grooves, which do have directionality. Leach specifies that $R_a$ measurements should ``take place perpendicular to the lay'', where the lay is the main feature direction or texture direction considered in this paper \cite{iso_4287,leach_fundamental_2010}. The plausability of determining feature direction from profiles and $R_a$ values is shown in figure \ref{fig:Rotation}, sweep directions parallel to the main feature direction result in lower $R_a$ values which may be easily observed, however random surface variation will mask this effect if few sweeps from from 0$^{\circ}$ to 90$^{\circ}$ are taken.

The surface roughness approximations in this paper are limited in scope, a greater range of starting topographies would increase the usefulness of the findings. The results from the simulated surfaces may approximate some real world cases well, for example machined surfaces, where distinct directional texture will be present, and random peak features are less common. In contrast the presence of random peaks and relative absence of distinct directional textures impacted the results from Earth surface samples. Unlike measurements taken using laboratory equipment, the theoretical measurements in this study are not limited in precision or accuracy, or by surface area or feature size. However, the number of data points considered was within the range of modern equipment, around 100,000 points per scan \cite{bruker_DektakXT}. In addition no filtering was applied to the surfaces and features at all scales were included in the calculations, for example filtering measurements using wavelet decomposition to assess short or long scale areal measurements \cite{jiang_paradigm_2007}. The use of areal and profile filters, not investigated in this work, introduces a fundamental difference between $R_a$ and $S_a$, choosing suitable cut-off values and aligning the rectangular areal measurement with the main surface features will help minimise differences \cite{leach_fundamental_2010}.

This work has only considered two comparable amplitude parameters: the arithmetic mean height deviation. Example parameters which have areal and profile equivalents are listed in table \ref{tab:parameters}. It should be noted that some areal parameters do not have a profile equivalent. The extent of agreement between additional $\bar{R}$ and $S$ parameters shown in table \ref{tab:parameters} will be investigated in future.

\Table{\label{tab:parameters} Example equivalent areal and profile parameters.}
\br
Profile & Areal & \\ 
Parameter & Parameter & Brief Description \\
\mr
$R_a$ & $S_a$ & Arithmetic mean height deviation  \\
$R_z$ & $S_z$ & Ten point mean roughness  \\
$R_q$ & $S_q$ & Root mean square (RMS) roughness  \\
$R_{sk}$ & $S_{sk}$ & Skewness  \\
$R_{ku}$ & $S_{ku}$ & Kurtosis  \\
\br
\endTable

This paper has shown that on surfaces with regular underlying features $\bar{R_{a}}$ approximates $S_a$ well as long as underlying feature direction is accounted for. Averaging between 3 and 5 $R_a$ values appears to be a reasonable compromise in time taken and accuracy, while very little improvement is observed when averaging \textgreater 10 $R_a$ values. In real world applications $\bar{R_{a}}$ approximates $S_a$ poorly, even when underlying feature direction is accounted for, however the results from ETOPO1 samples also show that little improvement is observed when averaging greater than 3 to 5 $R_a$ values. This paper does not cover the range of possible surface topographies. Extensions to the work should consider the effects of sample rotation on the $R_a$ values, and compare further $R$, $\bar{R}$ and $S$  parameter values for a range of surfaces. Caution should be taken extrapolating these outcomes to practical work without further generalisation of the model or more accurate models of surface roughness \cite{akande_modeling_2012,escobar_applications_2014,uchidate_generation_2011}.

\section{Supplementary materials}

The script used for the simulated surface part of this work is provided (Surface\textunderscore Creator.m), dependencies Ra.m and Sa.m are provided. Additional scripts which modify this to create charts or rotate surfaces are available from the author on request.

\ack 
HL thanks Anne Vanhoestenberghe (UCL) for helpful comments on a draft version of this manuscript.
HL wishes to acknowledge Engineering and Physical Sciences Research Council (UK), for financial support under their Centres for Doctoral Training scheme (grant EP/G036675/1), and under their Doctoral Prize Fellowship scheme. HL also wishes to acknowledge The Wellcome Trust for financial support (grant 106574/Z/14/Z).

\section*{References}
\bibliographystyle{unsrt} 
\bibliography{Surface_Roughness}
%
%
%
%
%
%
%

\end{document}